\documentstyle[12pt]{article}
\begin{document}

\begin{center}
\Large

\bf{A Growth Model for Porous Sedimentary Rocks}\\
\vskip 2cm
\normalsize
Sujata Tarafdar and Shashwati Roy
\footnote{E-mail : sujata@juphys.ernet.in}\\
\vskip 1cm
Condensed Matter Physics Research Centre\\
Physics Department\\
Jadavpur University \\
Calcutta-700032, INDIA.\\
\end{center}
\vskip 3cm
\noindent \bf{Abstract}\\
\normalsize
A growth model for porous sedimentary rocks is proposed, using a simple computer
simulation algorithm. We generate the structure by ballistic deposition of 
particles
with a bimodal size distribution. Porosity and specific surface area are 
calculated varying the proportion of small and larger particles. Permeability 
and 
it's variation with porosity are studied. The fractal nature of the pore 
space is also discussed.
\newpage
\section{Introduction}
\hskip .6cm
Study of natural growth processes through models and computer simulation
is useful and instructive. Successful computer models give an insight into 
how a simple sequence of steps governed by stochastic or other  algorithms 
may generate a structure with very specific characteristics. The computer 
generated 
system can then be used to study other properties of interest.

A striking example is the diffusion limited aggregation or DLA model. DLA and
it's variation - the diffusion limited cluster-cluster aggregation (DLCA) can 
be used to generate patterns resembling growth of colloidal aggregates, 
bacterial
colonies, corals, dielectric breakdown and many similar systems \cite{vic}.
Another field where such growth models have been applied is the formation of 
rough surfaces
 \cite{sta}.

 An area where such studies  are lacking is the formation process of porous 
sedimentary
 rocks. The study of sedimentary rocks is a very important field due to 
application in oil 
 exploration, ground water flow problems, spread of pollutants and other such
 cases.The structure of sedimentary rocks show certain characteristic features
 which make them distinct from other porous materials - such as foams or 
aerogel 
 \cite{sah}. However, there is still no simple model which simulates the growth 
 process of sedimentary rocks.

 The present work is an attempt in this direction. We use a simple ballistic 
deposition
model on a square or cubic lattice with a bimodal particle size distribution.
This generates a realistic rock structure with porosity varying according to
the particle size distribution.

The paper is organised as follows; in the next section we briefly discuss the 
principal
characteristics of sedimentary rock structure. In section 3 our model is 
presented and the 
details of the computer simulation are given in section 4. In the 5th section
we present the results of our study and finally discuss the success and 
drawbacks of 
our approach.

\section{Sedimentary rocks : Structure and properties}
\hskip .6cm
The sedimentary rocks originate from accumulation of small grains of sand or 
clay
often together with organic material \cite{pet,chi} .Sedimentation takes place 
through the
action of wind or water, and leads to a highly porous (50-80\%) unconsolidated 
agglomerate. The sedimentation process is followed by compaction and diagenesis
causing the unconsolidated mass to become a consolidated rock by flow of pore 
filling
 fluids accompanied by cementation, dissolution and other chemical processes. 
The resulting
 secondary porosity is usually less ,but sometimes greater than the primary
 porosity.

Sedimentary rocks are divided into classes such as sandstones or limstones 
depending upon their 
composition. Our model has been developed with the sandstone structure in mind
,however, it is quite general.
The principal characteristics of sedimentary rocks are as follows;
1. The solid phase and pore phase are both connected.
2. Porosities are usually in the range (5-25\%).
3. Many studies report a fractal structure of the pore phase \cite{sah}.
Adsorption studies report that the pore-solid interface is fractal, but this 
method
gives no information whether the system is a volume fractal as well. But 
several scattering
experiments indicate that the pore space is a volume fractal. 4. the 
electrical conductivity of brine filled porous sedimentary rocks and 
   their permeability are found to follow two well known emperical laws,
   referred to as respectively Archie's law and Darcy's law \cite{phtd}.

Archie's law is 
\begin{equation}
\sigma \propto \sigma_0.\phi_{eff}^m\label{eq1} 
\end{equation}
Here $\sigma$ is the effective conductivity of the electrolyte filled rock,
 $\sigma_0$
is the conductivity of the electrolyte, $m$ is a constant called the
cementation exponent. Usually $m$ has values between 1.3 and 4.0. For 
consolidated 
sandstones the value lies between 1.8 and 2.0. 
Normally $\phi_{eff}$ means porosity of the connected pore space, i.e. 
excluding the
isolated pores, however some authors give it a more restricted definition.

The permeability is defined by Darcy's law
\begin{equation}
J=(K/\mu )\nabla P\label{eq 2}
\end{equation}
Here $J$ is the volumetric flow rate, $K$ the permeability, $\mu$ the viscosity
of the fluid flowing through the porous medium and $\nabla P$ the pressure 
gradient.

The permeability depends not only on the porosity, but also on other details of
the pore structure \cite{sch}. Obviously, while a rock with no porosity will
have zero permeability, two samples with same porosity may have very different 
permeabilities. Other relevant parameters characterising the permeability have
been suggested as - the specific surface area, tortuosity and connectedness of
the pore space. Definitions of the last two quantities are not very clear.

It is expected that a realistic model of a sedimentary rock will give a 
satisfactory 
description of the above properties.

\section{The Model}
\hskip .6cm
Our work has two objectives -- the proposed model should simulate the process 
of 
rock formation in a realistic way, and the final structure formed should be
representative of a natural sedimentary rock.

The model has been developed in both two and three dimensional versions on 
a squre and cubic lattice respectively with unit spacing. Sand particles are 
dropped onto a substrate from a definite height from a random horizontal 
position. 
If all particles are squares of unit size (or cubes in 3 dimension), 
obviously the whole space is filled 
up after a sufficient number of particles are allowed to deposit. 
The upper boundary  may be highly uneven. If however, the particles to be 
deposited
include a fraction of particles of size larger than unity, we get a structure
with voids left randomly. The rough upper boundary is cut off.

We find that a fraction $F\sim 0.01$ of larger particles can generate a 
significant porosity 
(about 8\%), and the porosity increases with $F$. For $F=1$, we get the maximum
porosity permissible for a specific size and shape of the larger particle. The
variation in porosity with $F$ gives a quite good logarithmic fit.

Let us discuss the two dimensional system first. For the larger sized particle
we tried out squares of dimension $2 \times 2$ and alternatively rectangles of 
dimension 
$2\times 1$ with qualitatively similar results. Finally we confined our 
attention to
the elongated larger grain version only, because real sand grains are reported 
to be
ellipsoidal with the long axis approximately twice the shortest \cite{chi,pet}.
 The 
random structure generated is shown in the figures 1(a)-(c). We show three
different porosity ranges for three values of $F$.

In the 3-dimensional version we choose $X$ - $Y$ plane horizontal, and the 
$Z$- axis 
to be vertical. The smaller grains are cubes of unit side, and the larger are 
rods with dimension $2\times 1\times 1$. Here the larger particles are allowed 
to
settle ballistically with long axis parallel to either $X$ - or $Y$ - axis, 
with equal
probability. A modified version allowing preferred orientation is also being
tried.

This simple model simulates the action of gravity through vertical ballistic
deposition. For the present we have not included the effects of drift or
diffusion in the horizontal direction but this may be introduced in future work.
We have studied the following features of the porous structure generated :-
\begin{enumerate}
\item The variation of average porosity with $F$ (the fraction of large 
particles).
and the variation of porosity with sample size for a particular $F$.
\item The specific surface area of the pore-solid interface.
\item The fractal nature of the pore space generated and Archie's law.
\item The permeability variation with porosity and specific surface area has 
been
calculated using the Kozeny relation \cite{koz}.
\end{enumerate}

In the following section we give details of the computer simulation and the 
results obtained.

\section{Computer Simulation}
In our work the two-dimensional system has been generated upto size $600\times 
600$
and the three dimensional upto $150\times 150\times 150$. We first generate a 
linear (or 
square for three-dimensions) array of side $N_x$ which is solid i.e. all
sites are occupied. An uneven layer is generated on top of this, by filling
up certain sites randomly , other being left vacant. These two layers 
constitute the
substrate.

Now the growth process by deposition starts. The site for deposition is chosen 
randomly at a height $N_z$ above the substrate ($N_z=N_x+10$). The particle to
be dropped is of size I (small) or II (elongated) with a probability $1-F$ or 
$F$.
Whether the particle is of size I or II is decided by a random number 
generator. The particle descends as long as the sites immediately below it are 
empty. It stops on encountering a filled site.
Then the next particle drops. the process is continued until a square (or cube) 
of side $N_x$ is saturated, i.e. the porosity does not  change further on 
adding more particles.

The structure generated in two-dimension is shown in figs 1 (a-c). We repeat 
the 
whole process for a large number of times $N_{av}$. We have averaged over 
200-500
runs in two and 50-100 runs in three dimension.

\subsection{Porosity and Fractal Nature}
\hskip .6cm
After generating the system of the largest possible size, the average porosity 
is calculated as (the number of pores) /(total number of sites) in boxes of
gradually increasing size.

A typical porosity variation with size is shown in figure 2, here $F$ is 
constant.
A constant porosity indicates a homogenous structure without any fractal
characteristics. We find however, that the porosity first stabilises to
an almost constant value for $N_x\sim 200$ in two and $N_x\sim 100$ in three
dimensions. But for still larger sizes there is a slow but definite decrease
in porosity. This shows that though the structure is homogenous for small 
sizes a
fractality appears above a certain cutoff scale. This effect is more prominent 
for
low porosities (i.e. small $F$). Our boxes are always smaller than $N_x$ to 
eliminate any end effects in the structure generated. In the subsequent 
discussion
$P$ means the constant value of the porosity obtained, before it starts to fall.

We have estimated the fractal dimension using the relation
\begin{equation}
M(N)\sim N^{d_f}
\end{equation}
Here $M$ is the number of pores in a box of size $N$ and $d_f$ is the fractal
dimension of pore space. We find $d_f\geq 1.99$ in two and 2.99 in three 
dimensions
initially but it
deviates towards $d_f\approx 1.90$ in two and 2.89 in three dimensions as 
system 
size increases. The value of $d_f$
is lower for low porosities. Due to limited availability of computer facility
we have not yet been able to generate larger sizes so as to give a converging
value of the fractal dimension.

Porosity variation with $F$ is shown in figures 3 in three dimension.
The curve obtained gives a good logarithmic fit. The maximum value of $P$ (for 
$F=1$)
is 0.50 in two and 0.63 in three dimensions. Figure-4 shows $log M(N)$ plotted
against $log N$ for $F=0.005$.

\subsection{Specific Surface Area}
\hskip .6cm
Pore dependent Properties of porous rocks depend not only on porosity but also 
on other 
chacteristics of the pore structure, the easiest to calculate is the specific
surface area -- the interface area per unit mass (or volume).

We have calculated $S$ -- the average pore-solid interface area per unit 
volume, 
in the structure generated. $S$ plays a crucial role in permeability of the 
rock.

The variation of $S$ with $P$ from our model is shown in Figure 5. As expected 
$S$
first increases with $P$ upto $P\approx 0.6$, after that the pores become so 
large 
that many of the vacant sites do not contribute to the interface and $S$ 
decreases.
To get an idea of the connectivity of the pore structure we calculate $I=S/P$  
this gives the average number of interfaces exposed per each pore site. For
completely isolated pores the maximum value $I$ can have is 4 in two and 6 in
three dimension. $I\leq 4$ is a necessary but not sufficient condition for 
connectedness in three dimension. 
We have calculated $I$ only in three - dimension and find that $I$ remains 
close to 3, even for porosities near 4\%. This
indicates a high connectivity of the pore space. The variation of $I$ with $P$ 
is
shown in figure-6. At $P \approx 0.2\%$ $I$ approaches 6, so pores definitely
become isolated. The very sharp fall in $I$ between $P=14\%$ to $P=6\%$ seems to
indicate that the connectivity threshold is in this region, but this must be
verified.

\section{Results and Discussion}
\hskip .6cm
We now discuss the results obtained and their relevance to the features of rock
structures discussed earlier. $P$ for a given $F$ is taken as 
the constant value obtained for intermediate sizes (200 in two and 100 in three
dimensions).

In two dimension the maximum porosity for $F=1$ is 0.50. This is the percolation
threshold for a square lattice in the random percolation (RP) problem 
\cite{per}.
It is instructive to compare our model with RP. Here we have ensured the 
connectivity
of the solid phase through our growth algorithm and obviously it is not possible
to have both the pore and solid phases connected in two dimensions. But the 
diagrams generated in two dimension figs 1 show that a high degree of 
connectivity
in pore space is present for large $F$, particularly in the vertical direction,
 i.e. pores
are not isolated. There is an anisotropy in the structure generated in our
model with pores elongated in the vertical direction. This shows up more clearly
in the low porosity case. There is in fact some evidence of pores having higher
aspect ratio in case of low porosity \cite{asp}. There is also evidence of 
anisotropy of sedimentary rock structures, though there appears to be no
systematic study \cite{wie}.

The three dimensional structure our model generates, cannot be shown in a 
two-dimensional figure, and it is not possible to see the connectivity directly.
We have plans to make a complete study of the threshold porosity for 
connectedness
and the pore size distribution.The percolation threshold for RP in three 
dimensions
is found to be $P_c \approx 25-30\%$, and for porosities in the range $P_c$ to 
$(1-P_c)$ both pore and solid phases are connected.
it will be of interest to see whether we get a connected pore space here for 
lower porosities. 

Real rock structures are seen to be 
connected for very small porosities, in fact some authors speak of a zero
percolation threshold \cite{sah}. At present we have calculated the porosity,
specific surface area and permeabilty with interesting results.

\subsection{Permeability}

The permeability is an important property of a porous rock. It depends in a 
complicated manner on the structure of the pore space. $P$ and $S$ may be 
considered
as the simplest parameters characterizing a           pore space. To our 
knowledge
no exact theory connecting the permeability with $P$ and $S$ or any other such
parameter exists, for the case of such complicated multiply connected pore 
space as
in our model.

We may however get a qualitative idea of permeability ($K$) behavior using the 
oversimplified Kozeny relation \cite{koz} which gives $K$ as
\begin{equation}
K\sim {P^3}/{S^2}
\end{equation}
In the case of a pore space consisting of a bundle of unconnected tubes.

Usually application of the Kozeny relation considers $P$ and $S$ as separate 
independently varying parameters. In our model, however, they are related both 
being obtained simultaneously as a function of $F$.

We omit the tortuosity factor appearing in Kozeny relation as a proportionality
constant and consider only relative values of $K$ for different $F$ and plot it 
as a function of $P$. 

The results are shown in figure -7. We find a very good exponential fit. 
This implies that $$log(K)\propto  P$$ A large body of data for real
rocks do show such a variation \cite{chi}. For sandstones from different sources
log(K) vs porosity shows on the average a quite clear linear variation. There
is of course a considerable spread in the data for real rocks.

This may indicate that even in the case of such a complicated pore space, the 
Kozeny picture is not a very bad approximation as far as average flow properties
are concerned. In determination of $K$ the values of $S$ and $P$ play more 
crucial 
roles than the connectivity, which is missing in Kozeny relation. 
\subsection{Conductivity}

It has been shown that Archie's law (eq 1) connecting the conductivity of 
brine-filled
rocks with their porosity is valid for fractal pore spaces \cite{sas,kat,tho}.
Though we have not been able to conclusively demonstrate the fractal nature
of the pore space generated by our model, we can make some conjectures from
the results we have obtained.

For a fractal pore space, the cementation exponent depends on $d_f$ -- the
fractal dimension of the pore space and $d_w$ -- the dimension of a random
walk through the pore space as follows 
\begin{equation}
m=1+(2-d_w)/(d_f-d)
\end{equation}
For $d_f$ close to 3 in three dimension , say 2.85--2.90 as indicated by our 
simulation we may assume a value of $d_w$ slightly larger than 2, say 2.1--2.2.
With these values we get a range of m as 1.66 to 3, which is very realistic.

However, this results should be substantiated by an exact determination of 
$d_w$.
Work on this problem is in progress.

In comparison to  \cite{sas} and \cite{tho}, where the rock is
modelled as a deterministic fractal, the present model is more realistic being
a statistical model.

\section{Conclusions and Directions for Future Work}

Let us review how far our growth model represents actual formation of 
sandstones.
The first simplification is the assumption of cubic grains. In sanstones the 
grains
do in fact have somewhat angular form \cite{pet}, so this assumption  is not
worse than the more usual approximation of spherical grains. A consequence of 
the grains being cubic in our model is that the intergrain contact areas are
large and flat, whereas the sandstones are reported to have tangential point
contact between grains. This, however may take care partially of the compaction
after deposition in our model, which we have otherwise ignored.

We have not considered the effect of restructuring of the pore space, after 
deposition through diagenesis. This may be taken up in future work.
The elongated structure of the larger grains , which we have assumed, is
realistic as mentioned earlier. We have also checked that square flat grains 
produce a higher porosity than $P_{max}$ in our 3-dimensional model.

One point to note is that the porosity we have calculated from our model is
the total pore space per unit volume, not the effective porosity, which only
considers the connected pore space. Calculation of the effective porosity is 
much more complicated, and we assume for the present that since the pore
space is substantially connected the difference is not too much. In Archie's
law, some authors use the total porosity, and others the effective porosity.

Our calculation of pore space may also be modified by taking second nearest
neighbor pores as connected. This would give a picture of pores connected by 
narow throats as modelled by some authors \cite{sen}.

The structure produced by our simulation algorithm has an anisotropy in the
vertical direction. Pores are seen to elongated in the vertical direction, 
specially for low porosities. We could not find any systematic data on
anisotropy of pore structure in real rocks, but it is expected that due to
the action of gravity an anisotropy should be present. The permeability data
of some rocks show a marked anisotropy \cite{sch}, which indicates anisotropy 
in the
pore structure. The Kozeny theory does not take anisotropy into account since 
porosity
and specific surface area are independent of direction. Experimental data
shows the hprizontal conductivity to be larger than the vertical in most 
cases, but  the opposite trend is also seen sometimes.

In our model, connectivity appears to be larger in the vertical direction,
inclusion of diagenesis in some form may increase the horizontal connectivity
through flow of fluids.

Further modifications of this model are possible by including a horizontal
drift or diffusion to simulate the effect of wind or water currents on the
sedimentation process, in addition to the effect of gravity.

Comparing our model with the ballistic deposition model for surface growth
\cite{sta}, we note that our model introduces correlation between the adjacent
growing columns, through incorporation of the larger sized particle.

In conclusion, we think the present model is a convenient starting point
for study of the growth and structure of sedimentary rocks through simulation.
Suitable modifications and further study of certain aspects of the present
model are necessary, some of this work is already in progress.
\section{Acknowledgement}
S Roy is grateful to the State Govt., West Bengal for financial assistance.

\newpage
\noindent
\bf{Figure Captions}\\
\vskip 2cm
\noindent
{\bf Figure 1(a)--(c)} : Pore structure generated in 2-dimensions with (a) -- 
$F=0.2$,
(b) -- $F=0.4$, and (c) -- $F=1.0$. Open squares represent pore sites.
\vskip 1cm
\noindent
{\bf Figure 2(a)} : Porosity variation with sample size in 3-dimensional 
structure,
for $F=0.10$.
\vskip 1cm
\noindent
\bf{Figure 2(b)} : Porosity variation with sample size in 3-dimensions for 
$F=0.01$.
\vskip 1cm
\noindent
\bf{Figure 3} : Porosity variation with $F$ in 3-dimensions.Solid line shows 
logarithmic
fit, and broken line shows simulation results.
\vskip 1cm
\noindent
\bf{Figure 4} : Log[ M(N)] plotted against Log[N], the slope deviates from 3 
for large N.
\vskip 1cm
\noindent
\bf{Figure 5} :Specific surface area versus porosity in 3-dimensions.
\vskip 1cm
\noindent
\bf{Figure 6} :(Surface area)/(no. of pore sites) versus porosity.
\vskip 1cm
\noindent
\bf{Figure 7} : Relative permeability (K) versus porosity.
 Solid line shows exponential
fit, points are simulation results.  

\end{document}